**Mechanical Signature of a Chemically Driven Bath**

Tian Huang[1], Jintae Park[2], Hyuk Kyu Pak[2], Steve Granick[*,1]

[1]Department of Polymer Science and Engineering, University of Massachusetts Amherst, Amherst, MA 01003, USA
[2]Institute for Basic Science, Ulsan National Institute of Science and Technology, Ulsan 44919, South Korea

**ABSTRACT.** Using optical tweezers, we demonstrate that the CuAAC click reaction increases the positional fluctuations of a trapped 2 μm colloid by at least 20% without altering the trap's corner frequency. This excess variance relaxes to the thermal baseline as reagents are depleted. The additional noise is Gaussian, with a flat power spectrum extending past the corner frequency up to 7 kHz. Notably, the noise amplitude decouples from the macroscale reaction rate, and independent molecular force dipoles are orders of magnitude too small to generate the observed variance. Although the chemical free energy released locally exceeds the energy absorbed by the bead by five orders of magnitude, this energy couples inefficiently. The mechanical action of the reacting bath cannot be represented as a sum of independent molecular events.

Whether chemical turnover drives measurable departures from equilibrium fluctuations in the surrounding fluid, and whether the active-matter notion of athermal driving survives at molecular scales, remains open [1-6]. Evidence for catalysis-induced mobility has come almost entirely from single-molecule fluorescence, NMR, and specialized diffusion assays [7-13], each reporting a transport coefficient inferred from the motion of the reacting species themselves. Critics have argued that these inferences are confounded by photophysical artifacts, convection, transient heating, or concentration gradients [14-19]. Here we focus on the mechanical quantity at issue: the fluctuation power that chemical turnover delivers to a body in the fluid. Nonequilibrium fluctuations have been characterized in active baths [20], enzyme baths [21, 22] and living cells [23], but a mechanical fingerprint of a bath driven purely by non-enzymatic, small-molecule chemistry has not been established, and theory for how nanoscale chemical activity couples to larger bodies [24-26] has not been confronted with such a measurement.

In this Letter, we use optical tweezers [27] to probe the mechanical fluctuations of a passive probe in an aqueous catalytic 'click' reaction bath. The probe is much larger than the catalysts, so it coarse-grains molecular-scale chemical activity into measurable nonequilibrium noise. At equilibrium the bead's position variance is fixed by equipartition, $\langle x^2 \rangle = k_B T/k$, so only a rise in T or a fall in the trap stiffness k can mimic athermal forcing. We show that neither accounts for the data. The data we report requires none of the spectroscopic or photophysical assumptions that have fueled debate about enhanced molecular motility.

We trap 2 μm polystyrene spheres with a homebuilt optical tweezer [28] (1064 nm, 100x, NA = 1.4) in an aqueous CuAAC (copper-catalyzed azide-alkyne cycloaddition) reaction, chosen because it gave the largest enhancement in our earlier survey [12]. Preparation and filtration are given in [27]. Unless stated, $[CuSO_4]$ = 25 mM and ascorbate concentration is in four-fold excess, except where noted in the figure captions. The probe is held 25 μm above the coverslip, where the Faxén correction to the parallel drag coefficient (2%) is constant and cancels from active-to-baseline ratios. At 4 mW in the sample plane the trap stiffness is $k \approx 6$ pN/$\mu$m, giving a corner frequency $f_c = k/2\pi\gamma \approx 52$ Hz and probe relaxation time $\tau_{trap} = \gamma/k \approx 3.1$ ms with $\gamma = 6\pi\eta R = 1.9\times10^{-8}$ kg s$^{-1}$. Positions are recorded by quadrant photodiode at 14 kHz; displacement and force fluctuations are related through k. Measurements began at t ≈ 20 min, the minimum for mixing, cell assembly, and trapping. The signal is already decaying then, so our reported enhancements are lower bounds on their peak; their variance is elevated by at least 20% with $f_c$ constant.

During CuAAC turnover, the trapped probe exhibits a transient enhancement in fluctuation amplitude (**Fig. 1a**) while its one-dimensional displacement distribution P($x$) remains Gaussian (**Fig. 1b**) with an elevated variance relative to the equilibrium baseline. Power spectral analysis shows that the Lorentzian profile is preserved with the corner frequency unchanged throughout the reaction (**Fig. 1c**) and that the ratio of the active spectrum to the baseline spectrum is flat across the measured bandwidth (**Fig. 1d**). A viscosity change would change the plateau and the corner frequency $f_c$ while leaving the variance fixed. A drift in the detection sensitivity would change the variance and the plateau while leaving $f_c$ fixed. We observe the variance and the plateau elevated with $f_c$ constant, however.

The ratio in **Fig. 1d** is flat through $f_c$ and to the 7 kHz Nyquist limit of the 14 kHz acquisition, bounding the forcing correlation time at $\tau_c \lesssim 1/(2\pi \cdot 7\text{ kHz}) \approx 20$ μs; flatness through $f_c$ alone would bound $\tau_c$ only at the millisecond scale. The probe therefore integrates the forcing as white noise, averaging $\tau_{trap}/\tau_c \geq 150$ independent force samples per relaxation time, consistent with the Gaussian P(x). The active spectrum retains the $1/f^2$ roll-off with no 1/f or $1/f^3$ component of the kind produced by convection or long-range correlated forcing. Unlike bacterial baths [29-31], it shows neither ballistic excursions nor non-Gaussian tails.

*Contact author: sgranick@umass.edu, orcid.org/0000-0003-4775-2202

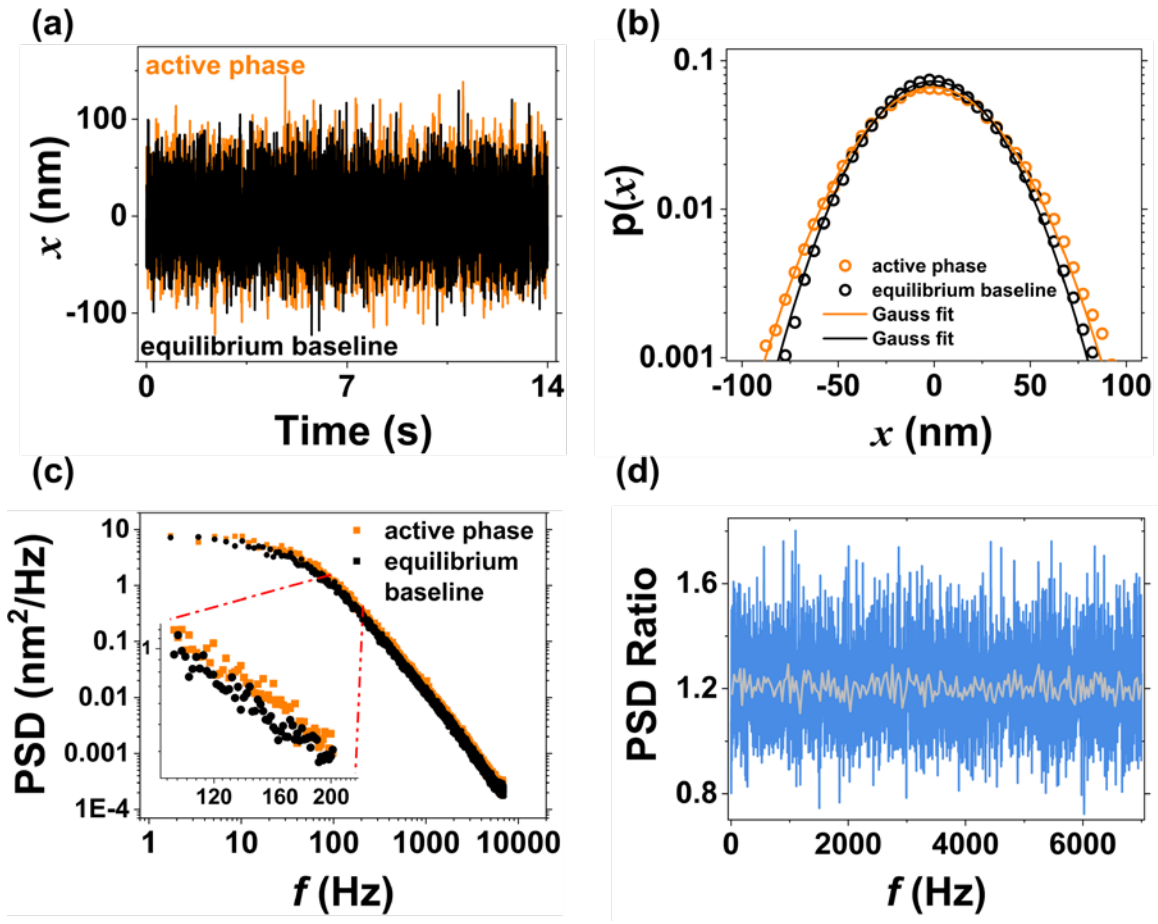

FIG. 1. Excess fluctuations in a CuAAC reactive bath at $[CuSO_4] = 25$ mM. **(a)** one-dimensional trajectories $x(t)$, **(b)** position probability density P(x), bin size 5 nm, with Gaussian fits, and **(c)** displacement power spectral density, each for the active phase ($t \approx$ 20 min after mixing, orange) and at the equilibrium baseline ($t \approx$ 120 min, black), from one sample. **(d)** Ratio of the two spectra in (c), point by point (blue) and smoothed over 20 consecutive points (grey). The corner frequency is unchanged between the two phases.

To rule out trivial origins, we also tracked bulk properties during turnover (**Fig. 2a-d**). The pH stays constant (**Fig. 2a**). Under equipartition, a 20% increase in variance at fixed k implies effective temperature $T_{\mathrm{eff}} \approx 1.2\ T$, a non-physical heating of $\Delta T \approx 60$ K, whereas thermometry in the trapping cell bounds $\Delta T < 0.3$ °C from the onset of trapping at $t$ = 20 min, decaying thereafter (**Fig. 2b**). Viscosity varies by <1% (**Fig. 2c**). The refractive index rises by 0.04% (**Fig. 2d**), which changes the polarizability factor $(m^2-1)/(m^2+2)$ by 0.2%, so $\delta k/k \lesssim 1\%$ and anyway the sign is opposite to what is needed, since variance falls while n rises monotonically.

Beads of different surface chemistry (unfunctionalized, carboxylated, aminated: zeta potential $\zeta = -35, -30$ and $+15$ mV in deionized water) give indistinguishable enhancement (**Fig. 2e**) and at our ionic strength ≈0.13 M the Debye length is $\lambda_D \lesssim 1$ nm suppresses electrophoretic mobility in any case. Phoresis does not explain this data; a steady phoretic slip [32] would displace the trap center rather than raise the variance, and a propulsive force decorrelates on the bead's rotational time, $4\pi\eta R^3/k_BT \approx 3$ s, so its spectrum rolls off above ≈0.05 Hz and does not contribute at our frequency range. Fluctuating gradients are bounded by the concentration-fluctuation estimate below.

Controls in **Fig. 2e** and **Fig. 3a** separate activity from composition. Without $CuSO_4$ no enhancement appears. That control does not evolve in composition, and Cu/ascorbate cannot be tested alone because Cu(I) precipitates without stabilizing ligands. At 10 mM $CuSO_4$ the reaction proceeds slowly, and by t = 140 min it approaches the composition present at the onset of the higher-catalyst runs (**Fig. S2a**), yet no enhancement is resolved, bounding a composition-only origin. The enhancement therefore requires ongoing reaction and is not set by composition alone.

We next compared the probe's excess fluctuations with molecular mobility in the bath. Because fluorescent markers were unstable at the low pH and water relaxation is long and variable, we used convection-compensated PFG (pulsed field gradient) NMR of oxidized ascorbate, which is formed during mixing when ascorbate reduces Cu(II) to catalytically active Cu(I). Its concentration is nearly constant, so its apparent diffusion coefficient $D_{app}$ serves as measurement of reaction progress. We do not address the contested matter whether $D_{app}$ also reflects a true mobility enhancement [14-19]; the mechanical findings we report do not rest on it.

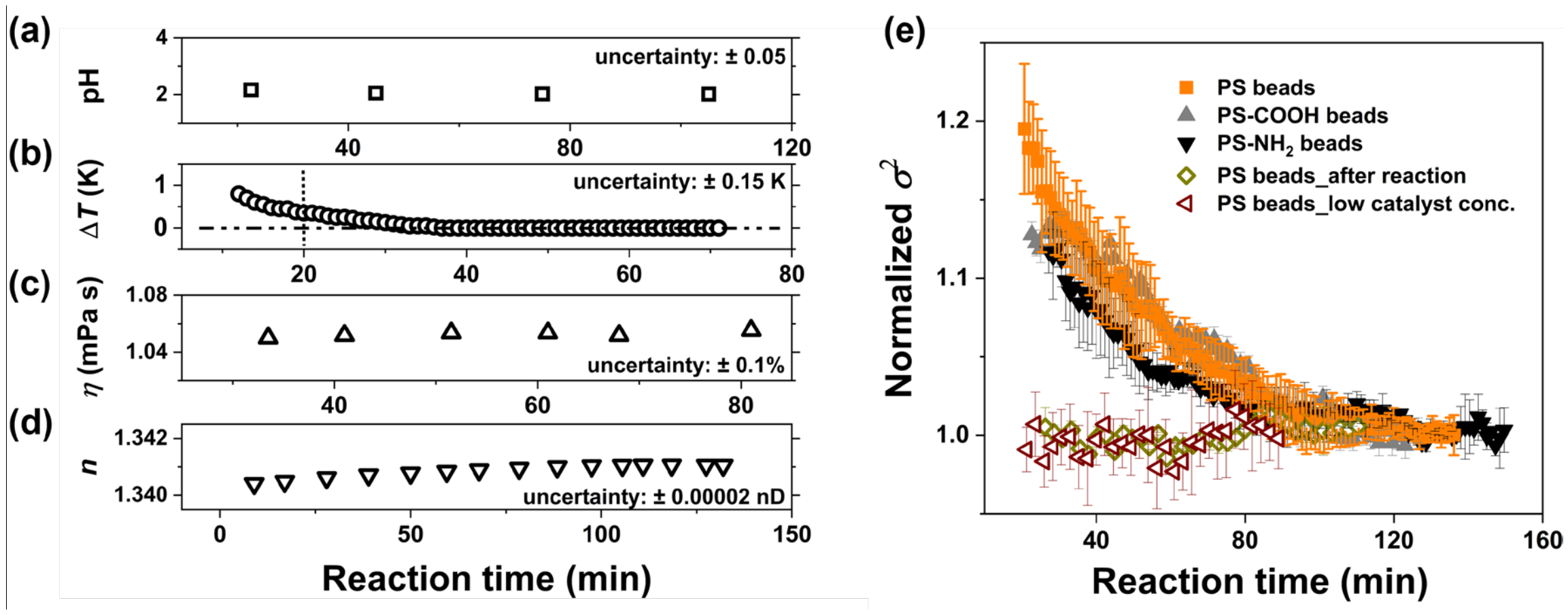

FIG 2. Bulk properties and control experiments. **(a)** pH, **(b)** temperature change in the trapping cell, **(c)** viscosity $\eta$, and **(d)** refractive index $n$ versus reaction time; error bars give the instrumental uncertainty; instruments are listed in [27]. Panels (a), (c), (d) were measured in bulk solution. **(e)** Normalized displacement variance $\sigma^2/\sigma^2_{\mathrm{final}}$ against reaction time for unfunctionalized, carboxylated, and aminated 2 µm polystyrene probes at $[CuSO_4]$ = 25 mM; 70 s sliding window average over seven independent runs per chemistry, bands are 95% confidence intervals. Diamonds (green): the same bead after reaction ends. Triangles (wine color): 10 mM $CuSO_4$ , the low-catalyst control.

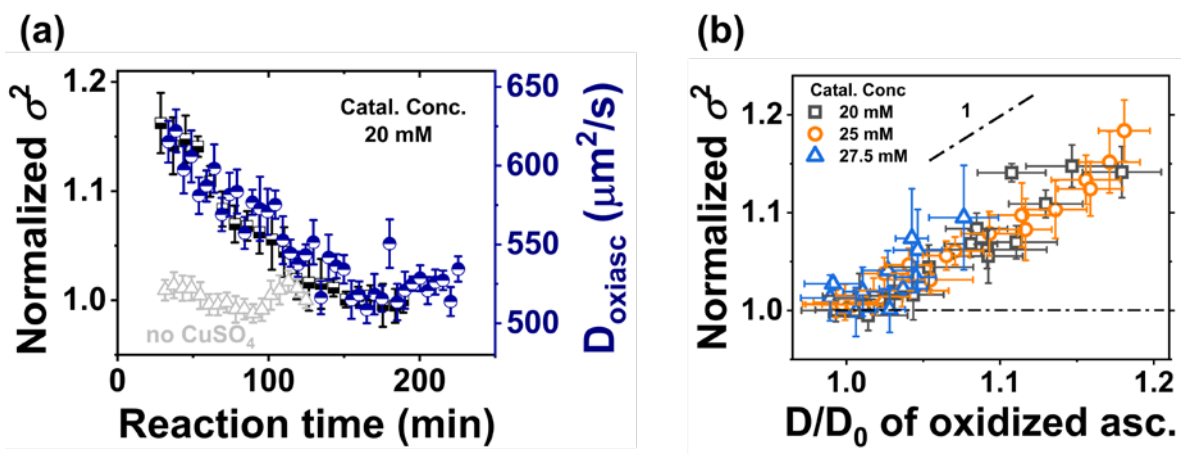

FIG. 3. Bead variance and molecular diffusivity track a common reaction coordinate. **(a)** Normalized variance $\sigma(t)^2/\sigma^2_{\mathrm{final}}$ (black, left axis) and apparent diffusion coefficient $D_{\mathrm{app}}$ of oxidized ascorbate (navy, right axis) versus reaction time at 20 mM $CuSO_4$ with 80 mM ascorbate. Triangles (grey): beads trapped at the same reagent concentrations without $CuSO_4$, the no-catalyst control. **(b)** $\sigma^2/\sigma_{final}l^2$ versus $D_{\mathrm{app}}/D_0$ at 20, 25, and 27.5 mM $CuSO_4$, each normalized to its post-reaction baseline; the dotted line marks unit slope as a guide to the eye. Spread along the curve reflects progress of the reaction, late times at lower left. The ≈20 min dead time is fixed, so faster reactions begin further along the curve: the 27.5 mM data miss its upper end. Means of three trials, unfunctionalized beads, error bars 95% confidence intervals.

The probe's variance and $D_{\mathrm{app}}$ share the same transient kinetics (**Fig. 3a**). Plotted against each other, data for three catalyst concentrations with rates different three-fold (**Fig. S2**) fall on one curve (**Fig. 3b**), so they track a common variable rather than elapsed time or catalyst concentration separately, but the scatter is too large to fit a slope. The excess decays while the reaction rate is still near its maximum (**Fig. S2a**), so the amplitude does not follow the instantaneous reaction rate, and the collapse in **Fig. 3b** shows that it does not follow the catalyst-set reaction rate either. A Poisson train of independent events gives $\dot{n}$ proportionality, the volumetric event rate, so this excludes them without reference to dipole strength or duty cycle and locates the decay in the medium. The estimates below are evaluated at t ≈ 20 min. Consistently, the diffusivities of different species in this reaction follow different time courses and only that of the azide tracks the reaction rate [33, 34].

Time resolved $^1$H NMR reveals reaction-synchronous shifts in the proton resonances of ascorbate (**Fig. 4a**) and the HDO solvent peak (**Fig. 4b**) while an internal reference remains fixed (**Fig. S3**). As in an earlier study of the CuAAC system by others [*17*], this is consistent with systematic changes in the local electronic environment and the hydrogen bonding network that track the reaction kinetics (**Fig. 4c**). Gravimetric density determination shows a monotonic increase in solution density (**Fig. 4d**) which, taken with the reported negative reaction and activation volumes for CuAAC [35, 36], suggests rapid local compaction and solvation-shell reorganization as reactants are converted to products. The mechanical enhancement decays to baseline while the densification persists, implying that the athermal forcing is controlled by ongoing reaction rather than by static properties of the product. Elevated fluctuations at unchanged viscosity, hence unchanged dissipation, violate the equilibrium fluctuation-dissipation relation, as expected when a chemical free-energy flux breaks detailed balance. Because the reaction evolves over $\approx 10^2$ min >> $\tau_{trap}$, the bath is quasi-steady on the scale of each measurement.

Osmotic forces do not account for the enhancement. A steady osmotic force would displace the trap center, and equilibrium solute fluctuations are already in the thermal baseline, so only a nonequilibrium excess can contribute. The probe volume $V = (4/3)\pi R^3$ at a total solute concentration of ≈0.7 M holds $N \approx 1.8\text{x}10^9$ molecules, with $N^{1/2} \approx 4\text{x}10^4$ refreshed by diffusion on $R^2/D \approx 1$ ms, comparable to $\tau_{trap}$. At the event rate $\dot{n}$ = (molar rate)$\cdot N_A \approx 2.4\times10^{22}$ m$^{-3}$ s$^{-1}$, about 300 reactions occur in this volume per $\tau_{trap}$, a Poisson fluctuation of ≈20. This is $\approx 5\text{x}10^{-4}$ of the equilibrium number fluctuation in amplitude, and $\sim 10^{-7}$ in variance, against the fractional variance excess of 0.2 we observe.

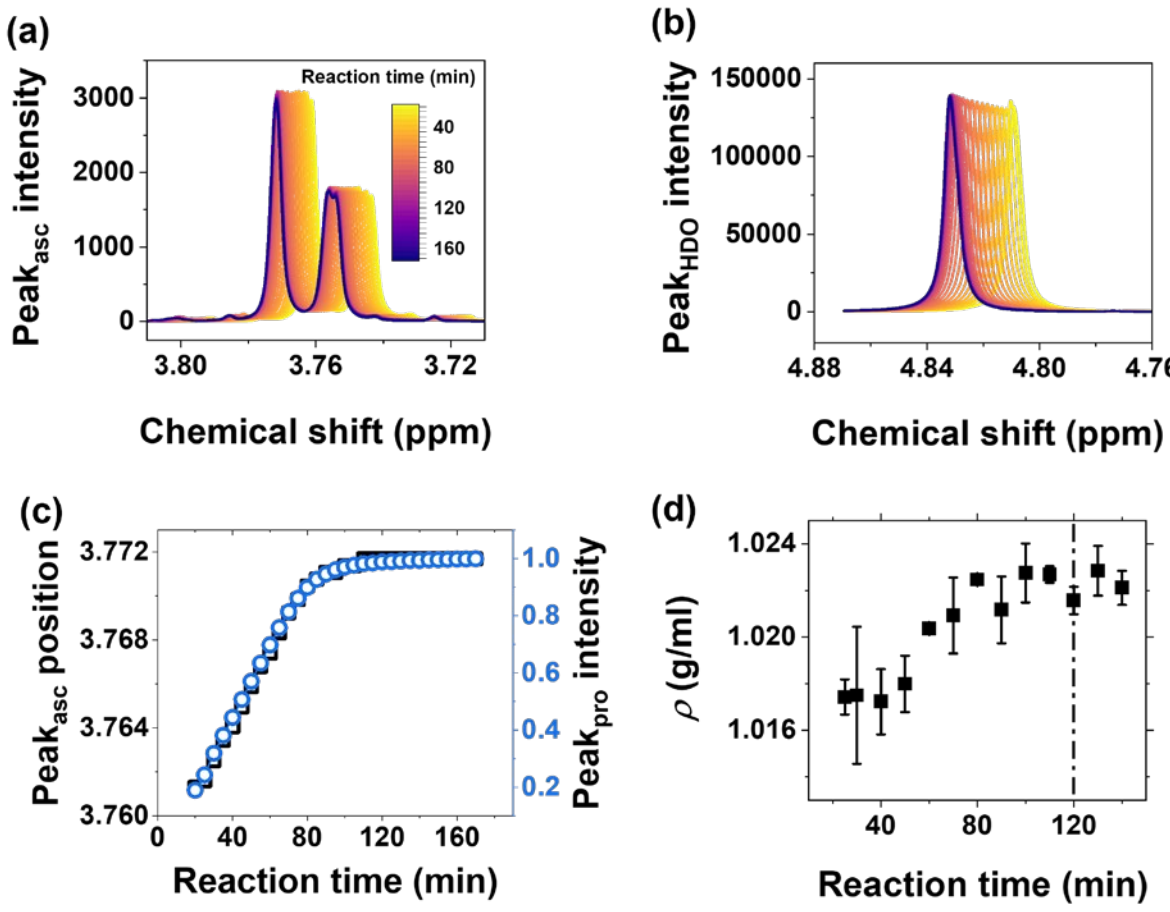

FIG. 4. Solvent reconfiguration tracks the reaction at $[CuSO_4]$ = 25 mM. Time resolved $^1$H NMR spectra of **(a)** the ascorbate and **(b)** the HDO resonances, colored by reaction time. **(c)** Ascorbate chemical shift (black squares, left axis) and reaction conversion from peak integration (blue circles, right axis) versus reaction time; the internal reference is unshifted over the same interval (Fig. S3). **(d)** Solution density ρ versus reaction time; the vertical line marks reaction completion at $t \approx 120$ min for this preparation.

A recent framework describes how nanoscale chemical activity generates stochastic hydrodynamic forces on larger probes [26]. Our observations are qualitatively consistent with it but at our dipole volume fraction and turnover cycle it predicts an enhancement some five orders of magnitude smaller than we measure [27]. These measurements place constraints on models of chemically driven baths. First, $\tau_c \leq 20$ μs. Second, the measured amplitude bounds the per-event forcing from below. Let us treat each reaction turnover as a transient force dipole of strength $\varepsilon|\Delta G|/N_A$ , with ΔG reaction free energy and ε the fraction converted to mechanical work. Then summing independent events at the measured rate $\dot{n} \approx 2.4\times10^{22}$ m$^{-3}$ s$^{-1}$, we find that the far-field flow reproduces the measured excess only for impulses lasting $\tau_e \approx (1/\varepsilon)30$ μs with the dipole axis staying fixed for the impulse duration. Third, this conflicts with the bound for any $\varepsilon \leq 0.3$ and is marginal even at the unphysical limit ε=1. A dipole in a ~1 nm reaction complex reorients in under a nanosecond, raising the discrepancy to a factor ≈330 in dipole moment [27]. Independent short-lived molecular events are therefore excluded.

The discrepancy stems from geometry, not energy availability; chemical free energy is released within the probe volume at $\approx 5\times10^{-14}$ W against the $\approx 3\times10^{-19}$ W the probe absorbs, a conversion of $\approx 6\times10^{-6}$ [27]. Because point-event flows decay as $1/r^2$ and M unaligned forces sum as $M^{1/2}$, overcoming these geometric losses requires the medium to correlate stresses across bead-scale regions of size ξ. Such a region is bounded below by holding one event per correlation time, $\xi \gtrsim (\dot{n}\tau_c)^{-1/3} \approx 1.3$ μm, and above by the momentum diffusion length $\sqrt{(\nu\tau_c)} \approx 4.5$ μm. But uniform aligned stress has zero divergence and drives no flow, so coherence recovers only $\approx 0.8\, \dot{n}R\tau_c\xi^2 \lesssim 10$ rather than the event count $M \approx 40$ [27], leaving the reorienting dipole short by $\gtrsim 10^4$ and rescuing the fixed-axis limit only for $\varepsilon \gtrsim 0.6$. Thus, no admissible correlation length supports a force-dipole description. The length is not measured here; two-point microrheology with dual traps could in principle supply it, whereas the probe-size route of bacterial baths [37] cannot, our range of trappable bead size being too short. A predictive theory must couple reaction–diffusion kinetics, hydrodynamics and nonthermal fluctuation spectra; until one exists, molecular interpretations of chemically driven fluctuations should remain cautious.

*Acknowledgments.* Experiments were initiated at the Institute for Basic Science, South Korea. Additional experiments, data analysis, and writing were supported by the grant DE-SC0025024 funded by the U.S. Department of Energy Office of Basic Sciences (TH, SG).

[1] C. Bechinger *et al.*, Active particles in complex and crowded environments. *Rev. Mod. Phys.* **88**, 045006 (2016).
[2] G. Gompper *et al.*, The 2025 motile active matter roadmap. *J. Phys. Condens. Matter* **37**, 143501 (2025).
[3] S. C. Takatori, T. Quah, J. B. Rawlings, Feedback Control of Active Matter. *Annu. Rev. Condens. Matter Phys.* **16**, 319-341 (2025).
[4] D. P. Arnold, A. Gubbala, S. C. Takatori, Active Surface Flows Accelerate the Coarsening of Lipid Membrane Domains. *Phys. Rev. Lett.* **131**, 128402 (2023).
[5] A. M. Tayar *et al.*, Controlling liquid–liquid phase behaviour with an active fluid. *Nat. Mater.* **22**, 1401-1408 (2023).
[6] O. E. Shklyaev, A. C. Balazs, Chemical signaling in reaction networks generates corresponding mechanical impulses. *PNAS Nexus* **4**, pgaf330 (2025).
[7] H. S. Muddana, S. Sengupta, T. E. Mallouk, A. Sen, P. J. Butler, Substrate Catalysis Enhances Single-Enzyme Diffusion. *J. Am. Chem. Soc.* **132**, 2110-2111 (2010).
[8] S. Sengupta *et al.*, Enzyme Molecules as Nanomotors. *J. Am. Chem. Soc.* **135**, 1406-1414 (2013).
[9] A.-Y. Jee, S. Dutta, Y.-K. Cho, T. Tlusty, S. Granick, Enzyme leaps fuel antichemotaxis. *Proc. Natl. Acad. Sci. U. S. A.* **115**, 14-18 (2018).
[10] A.-Y. Jee, Y.-K. Cho, S. Granick, T. Tlusty, Catalytic enzymes are active matter. *Proc. Natl. Acad. Sci. U. S. A.* **115**, E10812-E10821 (2018).
[11] K. K. Dey *et al.*, Dynamic Coupling at the Ångström Scale. *Angew. Chem. Int. Ed.* **55**, 1113-1117 (2016).
[12] H. Wang *et al.*, Boosted molecular mobility during common chemical reactions. *Science* **369**, 537-541 (2020).
[13] X. Zhao *et al.*, Substrate-driven chemotactic assembly in an enzyme cascade. *Nat. Chem.* **10**, 311-317 (2018).
[14] J.-P. Günther, M. Börsch, P. Fischer, Diffusion Measurements of Swimming Enzymes with Fluorescence Correlation Spectroscopy. *Acc. Chem. Res.* **51**, 1911-1920 (2018).
[15] I. Swan *et al.*, Sample convection in liquid-state NMR: Why it is always with us, and what we can do about it. *J. Magn. Reson.* **252**, 120-129 (2015).
[16] T. S. C. MacDonald, W. S. Price, R. D. Astumian, J. E. Beves, Enhanced Diffusion of Molecular Catalysts is Due to Convection. *Angew. Chem. Int. Ed.* **58**, 18864-18867 (2019).
[17] N. Rezaei-Ghaleh, J. Agudo-Canalejo, C. Griesinger, R. Golestanian, Molecular Diffusivity of Click Reaction Components: The Diffusion Enhancement Question. *J. Am. Chem. Soc.* **144**, 1380-1388 (2022).
[18] J.-P. Günther *et al.*, Comment on "Boosted molecular mobility during common chemical reactions". *Science* **371**, eabe8322 (2021).
[19] L. L. Fillbrook *et al.*, Following Molecular Mobility during Chemical Reactions: No Evidence for Active Propulsion. *J. Am. Chem. Soc.* **143**, 20884-20890 (2021).
[20] H. Seyforth, M. Gomez, W. B. Rogers, J. L. Ross, W. W. Ahmed, Nonequilibrium fluctuations and nonlinear response of an active bath. *Phys. Rev. Res.* **4**, 023043 (2022).
[21] M. Xu, J. L. Ross, L. Valdez, A. Sen, Direct Single Molecule Imaging of Enhanced Enzyme Diffusion. *Phys. Rev. Lett.* **123**, 128101 (2019).
[22] E. L. Mauricio Gomez *et al.*, Enhanced diffusion of colloidal tracers due to enzymatic activity. https://doi.org/10.48550/arXiv.2607.10646, (2026).
[23] R. Chakraborty, D. Paul, A. Maiti, K. R. Jayaprakash, K. K. Dey, Propagation of Enzyme-Driven Active Fluctuations in Crowded Milieu. *Small* **21**, e02935 (2025).
[24] A. S. Mikhailov, R. Kapral, Hydrodynamic collective effects of active protein machines in solution and lipid bilayers. *Proc. Natl. Acad. Sci. U. S. A.* **112**, E3639-E3644 (2015).
[25] J. M. Ortiz de Zárate, J. A. Fornés, J. V. Sengers, Long-wavelength nonequilibrium concentration fluctuations induced by the Soret effect. *Phys. Rev. E* **74**, 046305 (2006).
[26] A. K. Tripathi, T. Tlusty, Gauging Nanoswimmer Dynamics via the Motion of Large Bodies. *Phys. Rev. Lett.* **129**, 254502 (2022).
[27] See Supplemental Material at [URL link] for experimental methods, analysis of the results.
[28] J. T. Park, G. Paneru, C. Kwon, S. Granick, H. K. Pak, Rapid-prototyping a Brownian particle in an active bath. *Soft Matter* **16**, 8122-8127 (2020).
[29] X.-L. Wu, A. Libchaber, Particle Diffusion in a Quasi-Two-Dimensional Bacterial Bath. *Phys. Rev. Lett.* **84**, 3017-3020 (2000).
[30] D. T. N. Chen *et al.*, Fluctuations and Rheology in Active Bacterial Suspensions. *Phys. Rev. Lett.* **99**, 148302 (2007).
[31] E. Lauga, T. R. Powers, The hydrodynamics of swimming microorganisms. *Rep. Prog. Phys.* **72**, 096601 (2009).
[32] J. L. Anderson, Colloid Transport by Interfacial Forces. *Annu. Rev. Fluid Mech.* **21**, 61-99 (1989).
[33] T. Huang, B. Li, H. Wang, S. Granick, Molecules, the Ultimate Nanomotor: Linking Chemical Reaction Intermediates to their Molecular Diffusivity. *ACS Nano* **15**, 14947-14953 (2021).
[34] T. Huang, B. Li, H. Wang, S. Granick, Reply to "Comment on 'Molecules, the Ultimate Nanomotor: Linking Chemical Reaction Intermediates to their

Molecular Diffusivity'". *ACS Nano* **16**, 9977-9981 (2022).
[35] R. Van Eldik, T. Asano, W. J. Le Noble, Activation and reaction volumes in solution. 2. *Chem. Rev.* **89**, 549-688 (1989).
[36] H. Chen, B.-B. Ni, F. Gao, Y. Ma, Pressure-accelerated copper-free cycloaddition of azide and alkyne groups pre-organized in the crystalline state at room temperature. *Green Chem.* **14**, 2703-2705 (2012).
[37] A. E. Patteson, A. Gopinath, P. K. Purohit, P. E. Arratia, Particle diffusion in active fluids is non-monotonic in size. *Soft Matter* **12**, 2365-2372 (2016).